\documentclass[aps,prl,twocolumn,groupedaddress,showpacs,floatfix,color]{revtex4}
\usepackage{graphicx}
\usepackage{epsfig}
\usepackage{amsfonts}
\usepackage{color}
\usepackage{ulem}

\begin{document}

\title{Income and Poverty in a Developing Economy}

\author{
Amit K Chattopadhyay}
\affiliation{SUPA, School of Physics and Astronomy, The University of Edinburgh, Mayfield Road, Edinburgh, EH9 3JZ, UK.}

\affiliation{Department of Physics and Astrophysics, University of Delhi, 
New Delhi 110007, India.}

\author{Graeme J Ackland }
\affiliation{SUPA, School of Physics and Astronomy, The University of Edinburgh, Mayfield Road, Edinburgh, EH9 3JZ, UK.}
\email[Corresponding author: ] {gjackland@ed.ac.uk}

\author{Sushanta K Mallick}
\affiliation{School of Business and Management, Queen Mary
 University of London,
Mile End Road, London E1 4NS, UK}

\begin{abstract}
We present a stochastic agent-based model for the distribution of
personal incomes in a developing economy. We start with the assumption
that incomes are determined both by individual labour and by
stochastic effects of trading and investment. The income from personal
effort alone is distributed about a mean, while the income from trade,
which may be positive or negative, is proportional to the trader's
income. These assumptions lead to a Langevin model with multiplicative
noise, from which we derive a Fokker-Planck (FP) equation for the
income probability density function (IPDF) and its variation in
time. We find that high earners have a power-law income distribution
while the low income groups have a Levy IPDF.  Comparing our analysis
with the Indian survey data (obtained from the world bank website) 
taken over many years we obtain a near-perfect
data collapse onto our model's equilibrium IPDF. The theory
quantifies the economic notion of ``given other things''. Using survey data 
to relate the IPDF to actual food consumption we
define a poverty index \cite{sen1976,kakwani1980}, which is consistent
with traditional indices, but independent of an arbitrarily chosen
``poverty line'' and therefore less susceptible to manipulation.

\end{abstract}
\date{\today}
\maketitle

Poverty has been a feature of all human societies throughout time.
The underlying cause is the unequal distribution of
personal incomes which is an emergent feature of a free economy,
invariably resulting in extreme wealth for a few and relative poverty
for many.

Since the work of Pareto \cite{Mandelbrot}, the distribution
of incomes has been known to have a power law tail at the high
end \cite{Mantegna}.
There have been many models of the dynamic
process \cite{Mandelbrot,Montroll,bouchaud,book,Mohanty,chat} by which a
power-law tail can develop for high incomes.  Yet in an interlinked economy the low income
distribution emerges from the same dynamics as the high income.

Significantly less effort has been applied to study the distribution of
low-incomes, but to study poverty this is the critical part of the IPDF.  
Empirical data shows that low-income
distributions are not well described by a Pareto-style power law with
a sharp cutoff, as is typically introduced to obtain a normalizable
IPDF. Rather than curve-fitting to data, we seek to model the most
elementary processes of economic activity, and to find the
distribution which emerges.

Such interacting systems are well described using methods in 
statistical physics \cite{challet,johnson,During}.
Our basic idea is to represent each individual as an ``agent'',
generating income through personal labour and trade.  We describe
the income above starvation level, $y_i(t)$, of each agent, $i$,
with a stochastic dynamical equation which describes both labour
and trade.  It will turn out that trade is the crucial feature for
high income groups, while labour is important for lower income
groups.

We postulate that the time variation of agent income has the form of a Langevin equation:
\begin{equation}
\frac{dy_i}{dt} = C(t) - My_i + \eta_i(t)y_i
\label{1}
\end{equation}
Where $C$ represents the rate of increase in income possible from
labour, $My$ represents the increasing difficulty to maintain a high
income, and $\eta_i(t)$, a random variable with zero mean, represents
the stochastic effects of trading. $C(t)$ is a property of the economy
as a whole and is slowly varying in time. Possible gains from
employment depend on how the economy as a whole is performing. $M$ is
a constant which we shall later determine from empirical data. Note
that a non-zero mean for the noise term would be equivalent to a smaller
value of $M$, so no assumption is being made about net benefits of trade.

It can be seen that income from labour alone is the same for each person,
however the value of trading is proportional to an individual's
current wealth.  This mix introduces multiplicative noise which 
is in contrast with previous dynamical
approaches \cite{Cont,takayusu,Richmond,Picozzi} in economics, producing
anomalous diffusion from the noise itself, not fractional
dynamics \cite{Picozzi,Metzler}.
Equation 1 does not map on to any well-known physical
system, however, there is increasing neurological evidence for such
non-linear risk taking \cite{Glimcher,Delgado}.


While it is easy to postulate reasonable-looking intuitive theories
for income distribution, there are no known fundamental laws, and so
empirical verification is essential \cite{bouchaud2008}. The largest 
dataset available for personal income in
a developing economy is that collected by the Indian National Sample
Survey Organisation (NSSO) covering incomes of millions of people for almost 40 years \cite{NSSO}. 
 The same survey reveals the
fraction of income spent on staple food (cereals).  Since food is
the absolute minimum necessity for survival, we will base our measure of
poverty on expenses related to its consumption.

The raw NSSO data comprises income bands (``expenditure classes'') of
irregular size, from which we generate cumulative income
distribution functions (CDFs). Figure 1 shows three typical graphs 
out of 21 surveys across more than a million
households (household size varies from 4-6) between the
years 1959-1991.  Once scaled to
match the mean income for each year, there is a remarkable data
collapse. 
The inset shows that the IPDF emerging from our model is also 
in excellent approximation to this functional form, as we now discuss.

We assume that the trading decisions are made before
their outcome is known, which indicates that
we should use Ito calculus: had we assumed mid-term review of trading
strategy it would imply Stratonovich calculus, which leads to an
equivalent equation with rescaled $M$. This 
leads to a Fokker-Planck equation derived
from the Langevin  model (eq \ref{1}).
\begin{equation}
\frac{\partial {\hat f}}{\partial t}(y,t) =  \frac{\partial}{\partial y}
\:\{[(M+2)y - C(t)] \:{\hat f} + y^2\:\frac{\partial {\hat f}}{\partial y}\} 
\nonumber 
\label{2}
\end{equation}

In the steady state ($C(t)=C_0$), this would give us
the income distribution:
\begin{equation}
{\hat f(y,t\rightarrow \infty)}=
\frac{{C_0}^{M+1}}{\Gamma{(M+1)}}\;\frac{\exp(-C_0/y)}{y^{M+2}}
\label{3}
\end{equation}
\begin{center}
\begin{figure}[tbp]
\includegraphics[width=7cm]{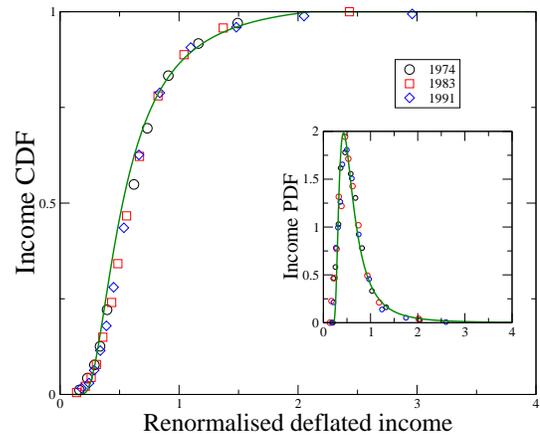}
\caption{Plots of the cumulative distribution functions (CDFs)
against
deflated income for selected years, with inflation independently
sourced from the consumer price index (CPI) and renormalized to the
1974 mean income in rupees (64.84 INR). 
The green line is our theoretical curve, taking $y_i$ as income 
above a non-zero level below which agents would die of starvation 
(set at 0.15 in renormalized units).  Inset shows the
IPDF which is the differential of the CDF, evaluated from
the data by interpolation.  The points are the data from the NSSO, the
lines are our analytic function for the steady state distribution, the
only fitting parameters are the power-law
tail exponent $M+2=3.6$
\label{fig_cdf}}
\end{figure}
\end{center}

Eigenvalue analysis shows that this solution is stable against
perturbations. Equation \ref{2} can be analytically solved using
Laplace transforms to obtain the full time-dependent solution as a sum 
of confluent hypergeometric functions $F(a,b,z)$ 
with time-dependent coefficients:

\begin{equation}
{\hat f}(y,t) = \sum_{n=0}^{n=\infty}\:\exp(-\omega_n t)\:g_n(y) 
\end{equation}

where  $\omega_n=2\pi n$ and

\begin{eqnarray}
g_n(y) &=& A_1\:{\left ( \frac{c(t)}{y} \right )}^
{\alpha_{-}}\: F(\alpha_{-},\beta_{-},-\frac{c(t)}{y}) \nonumber \\
&+& A_2\:{\left ( \frac{c(t)}{y} \right )}^
{\alpha_{+}}\: F(\alpha_{+},\beta_{+},-\frac{c(t)}{y})  
\\
\alpha_{\pm}&=&\frac{3+M \pm \sqrt{{(1+M)}^2+4\omega_n}}{2}
\\\beta_{\pm}&=&1 \pm \sqrt{{(1+M)}^2+4\omega_n}
\end{eqnarray}
and $A_1$ and $A_2$ are constants
dependent on initial conditions.

Any redistribution measure can be represented by a sum of these
functions, and from the associated decay times the timescale of its
effect can be determined.     However, the
data collapse in figure 1 suggests that the relaxation time back to
the steady state distribution is rapid.

There is one free parameter, $M$, which incorporates difficulty of 
maintaining high income, mean benefit of trade and any possible misconception
from our choice of Ito calculus. 
$M=1.6$ gives in extremely good agreement
with the NSSO data, with $M+2=3.6$ describing the exponent in the power law
tail, $C_0 \frac{\Gamma(M)}{\Gamma(M+1)}$ 
being the mean income. This gives us great confidence
that our simple, intuitive model does indeed describe the coarse
features of the Indian economy,
and allows us to proceed with our model of poverty.  

Whether poverty-reduction measures are regarded as successful
or not often depends on the precise definition of poverty, a semantic
which is still argued over.
Sen has defined the so-called ``axioms of poverty'' \cite{sen1976}.
\begin{itemize}
\item Given other things, a reduction in income of a person below the poverty line must increase poverty.
\item Given other things, a pure transfer of income from a person below the poverty line to anyone who is richer must increase poverty.
\end{itemize}
While these seem to be self-evident, they are based on the ill defined
notion of "other things'' being unchanged.  This is troublesome: it
implies that we should be dealing with partial derivative, but does
not specify exactly which variables should be held constant it is not
possible simply to increase mean income and "hold everything else
constant".  Worse, a dynamic system {\it will} have some non-trivial 
response to any income reduction of transfer.

By defining the process underlying the income distribution, it becomes
possible to define precisely what ``given other things'' means.  For example,
increasing mean income corresponds to increasing $C$ in equation
1.  A flat-rate tax: increasing low incomes, reducing high incomes,
maintaining mean income, would correspond to a multiplier on the
$C-My$ term, more complex tax arrangements would change its form
altogether.  The effect of such changes will induce both a transient 
response and a steady-state change in IPDF.

In the context of India, there has been considerable debate on the
`true' measure of poverty as the so-called ``poverty line'' estimates remain
controversial \cite{poverty_line}
yet crucial in the empirical literature on poverty
analysis. The official poverty lines which guide Planning Commission
policy are based on the nutritional need for calories, but these have
been criticized for under-estimating `true' poverty in India \cite{palmer2003}.

Three conventional poverty measures involve defining a certain income as the ``poverty line'', and counting 

(i) the fraction of the population with incomes below it (headcount index, HCI)

(ii) The mean percentage below the poverty line (poverty gap index, PG)

(iii)  The mean percentage squared below the poverty line (squared 
poverty gap index, SPG)

A difficulty with such measures is to define the ``poverty line'', a
somewhat arbitrary level of income which also changes with time due to
inflation.  The successive definitions of poverty measures above
reduce the sensitivity of the poverty index to this choice, but do not
eliminate it, and pathological cases can easily be derived, especially
in practice where NSSO data is discretized into expenditure
classes rather than a continuous.

To define a more robust poverty measure, we apply the idea of
consumption deprivation (CD) for a specific resource
\cite{kumar1996,sitaraman1996,pradhan2000}. This uses the fact that
expenditure on cereals is monotonically increasing with income, but
flattens above a certain income, reflecting the saturation of demand
for cereal once one has sufficient to avoid malnutrition.

Correlating the NSSO income data with that for cereal
expenditure, we find a good fit to a Monod relationship
\begin{equation}CD(y)= VK/(K+y)\label{Monod}\end{equation}
where the parameters $V$ and $K$ are time dependent \cite{kumar1996}. 
 Broadly, $K$ can be taken as a ``poverty
line'' which accounts indirectly for cereal-price inflation as opposed
to general inflation. $V$ measures the deflated price of cereals. A
more intuitive measure of deprivation is the quantity $CD(y)/V$, which
is the fraction of the maximum desirable cereal consumption actually
consumed by someone of income $y$.

The advantage of this measure is that it is based on people's actual
choices, not on the price of an arbitrarily chosen ``basket of
goods''. So, for example, increasing housing or clothing costs may
affect $CD$ even when cereal prices are steady, as income has to be
moved from one commodity to another to balance the overall budget.
Similarly $CD$ is not affected by changes in the
CPI due to price shifts of luxury goods
purchased only by the wealthy.  

Perhaps most importantly,  cereal consumption is
directly measured by the NSSO.
This allows us to assign a level of poverty to each such NSSO ``expenditure classes''. 
By summing this measure, a
poverty index based on actual consumption deprivation may be evaluated. We
refer to this as the $CD$-index of poverty, $P_{CD}$.

\begin{center}
\begin{figure}[tbp]
\includegraphics[width=7.0cm]{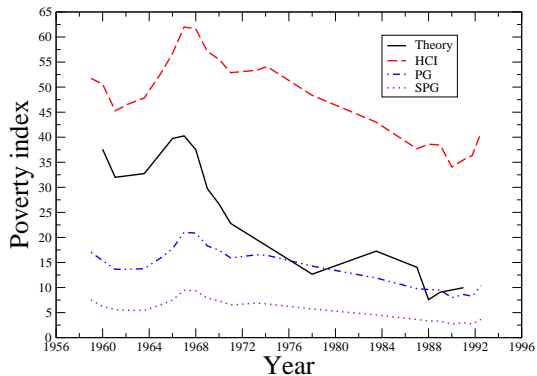}
\caption{Plots of the CD-index, a measure of poverty \cite{kumar1996},
against time. 
 The broken line shows the CD-index as 
per the official head-count poverty  index 
while the continuous line shows the poverty index arising from our
IPDF.  To generate the latter we have taken the function $C(t)$ in
equation 1 to be a piecewise-linear interpolation of the NSSO-measured
mean income at each round and assumed a relaxation time less than a
year.  Also shown are the poverty gap and squared poverty gap indices, 
with the poverty line set at the official rate 356 Indian rupees per month.
The absolute numbers cannot be compared, but the similarity in trends
is evident. $V(t)$ and $K(t)$ are defined from the NSSO data: graphs of
V(t), K(t) and C(t) are given in the Supplementary Material.
\label{fig_povI}}
\end{figure}
\end{center}

Our model allows us to quantify this $CD$-index of poverty. 
The model definition of the $CD$-index satisfies the standard
axioms of a poverty index \cite{kakwani1980,sen1976}, eliminates the arbitrary ``poverty line'', {\it and}
makes explicit the meaning of ``given other things''.  
Using the NSSO data, we can fit an analytic form \cite{kumar1996}
to the ratio of grain expenditure to income. The CD-index is then defined by
the relation

\begin{equation} 
P_{CD}(t) = \int \frac{V(t)K(t)}{K(t)+y} \hat{f}(y,t) dy
\end{equation} 

where parameters $V,\;K$ are obtained from NSSO data while
$\hat{f}(y,t)$ is the solution of equation \ref{2}.  The income data
used to parameterize our model is independent of the consumption data
used to measure $CD$ directly. In Fig. \ref{fig_povI}, we compare
the $P_{CD}(t)$ evaluated directly from the NSSO {\it consumption}
data, and indirectly from our {\it income-data} based model.  We also
show the PG and SPG indices.  All indices show poverty declining in
time, with a peak due to sharp drops in income in the 1960s.  However the CD-index shows the effect of increasing cereal
prices between 1978-84 as causing an increase in poverty, an effect which
cannot be captured in the standard indices.

Against the CPI-deflated data, we see that mean incomes have generally
risen over the last forty years, while the relative price of cereals
($V(t)$) has dropped steadily (see Supplementary Material). 
This helps to reduce poverty,
although more direct targetting \cite{Mexico} may be even more effective.


Returning to the stationary IPDF, the power law exponent M is seen to be
a crucial component in quantifying the mean income: 
$C_0\Gamma(M)/{\Gamma(M+1)}$. 
Critically, since we have shown that if trading is, on average,
beneficial rather than neutral, it will reduce M.
Small $M$ increases both the mean income and the level of inequality - 
it transfers capital from lower to
higher income groups. 

This illustrates a problem with Sen's axioms. Raising
mean incomes "given other things" reduces poverty, while transferring
income to higher income groups "given other things" increases
poverty. So, in this worldview, the effect of beneficial trade on
poverty depends on the definition of ``other things''.  
Although one can devise pathological cases, what we find
here is that the effect of increasing trade ($M\approx 1$) is to
reduce absolute poverty provided the mean income is above the
``poverty line'' for a headcount index or $K$ for $P_{CD}$.  However, it
also has the effect of increasing measures of ``relative poverty''
where the ``poverty line'' is a fixed fraction of the mean income.

In summary, we have postulated a stochastic model for the evolution
of the income distribution in a developing economy. 
The steady state of the distribution is
stable and robust, and in excellent agreement with the massive NSSO 
data set for Indian incomes over many years. The existence
of an underlying probability distribution function parameterized by
mean income makes it much easier to estimate poverty than existing
measures such as the head-count index. Under this measure the poverty
index is completely specified by the data, without recourse to
defining a ``poverty line''.  Moreover, the measure is less
susceptible to manipulation by distortions to the income distribution
around the poverty line: ``lifting people out of poverty'' (just).
A major strength of this theory lies in its potential
power of predicting the response of the IPDF, 
and hence the poverty index, to external effects, 
up to a reasonably close (perturbative) time span.

GJA and AKC would like to thank EPSRC for support of the NANIA 
project and discussions with J.Tailleur. AKC acknowledges helpful 
discussions with G. Rowlands, 
University of Warwick and R. Metzler, Technical University of Munich.

\end{document}